# Vertical-external-cavity surface-emitting lasers and quantum dot lasers

Guangcun SHAN (✉)[1, 2,*], Xinghai ZHAO[1, 3], Mingjun HU[2], Chan-Hung SHEK[2], Wei HUANG[4]

1 State Key Laboratory of Functional Materials for Informatics, Chinese Academy of Sciences, Shanghai 200050, China
2 Department of Physics and Materials Science, City University of Hong Kong, Hong Kong SAR, China
3 Institute of Electronic Engineering, China Academy of Engineering Physics, Mianyang 621900, China
4 Institute of Advanced Materials (IAM), Nanjing University of Posts and Telecommunications, Nanjing 210046, China

*To whom correspondence should be addressed.

Fax:+852-34420538

Email: guangcunshan@mail.sim.ac.cn.

**Abstract** The use of cavity to manipulate photon emission of quantum dots (QDs) has been opening unprecedented opportunities for realizing quantum functional nanophotonic devices and also quantum information devices. In particular, in the field of semiconductor lasers, QDs were introduced as a superior alternative to quantum wells to suppress the temperature dependence of the threshold current in vertical-external-cavity surface-emitting lasers (VECSELs). In this work, a review of properties and development of semiconductor VECSEL devices and QD laser devices is given. Based on the features of VECSEL devices, the main emphasis is put on the recent development of technological approach on semiconductor QD VECSELs. Then, from the viewpoint of both single QD nanolaser and cavity quantum electrodynamics (QED), a single-QD-cavity system resulting from the strong coupling of QD cavity is presented. A difference of this review from the other existing works on semiconductor VECSEL devices is that we will cover both the fundamental aspects and technological approaches of QD VECSEL devices. And lastly, the presented review here has provided a deep insight into useful guideline for the development of QD VECSEL technology and future quantum functional nanophotonic devices and monolithic photonic integrated circuits (MPhICs).

## 1 Introduction

Semiconductor quantum dots (QDs) have attracted a lot of interest for both their fundamental physics and potential applications ranging from optoelectronics to quantum information processing [1–9].

---

*E-mail: guangcunshan@mail.sim.ac.cn

Epitaxially grown semiconductor QDs by the Stranski-Krastanov self-organized method using molecular beam epitaxy (MBE) or metalorganic vapor phase epitaxy (MOVPE) have demonstrated their potential as ultrafast semiconductor lasers [5,6] and semiconductor optical amplifiers (SOAs)[7] for optical communication systems, showing new functionalities and excellent performances. Notably, a different approach for the fabrication of QDs is based on wet chemistry for synthesizing of colloidal QDs [8], and a facile microwave-assisted synthesis of the core-shell NCs in aqueous solution has been further developed in 2006, which opened up a promising avenue for synthesizing NCs with high fluorescence efficiency [9]. Typical colloidal QDs are nanometer-sized core-shell structures, and the role of the shell is to engineer the band structure of the QDs and to passivate the core surface to reduce surface defects and improve their efficiency and photostability [9,10].

On the other hand, semiconductor lasers are used in a wide range of important applications, such as optical fiber communication, digital optical recording, laser materials processing, biology and medicine, imaging, spectroscopy, and some others. In particular, the vertical-external-cavity surface-emitting lasers (VECSELs) developed in the mid-1990s [11] to overcome problems with conventional semiconductor lasers, have gained a reputation as a superior technology for optical communication system applications. This success was mainly due to the VECSEL's lower integration manufacturing costs and higher reliability compared to semiconductor edge-emitting lasers. Moreover, recent years have seen an increasing use of semiconductor QDs as gain component of different VECSEL devices of design [6, 12, 13]. In this review, we discuss the properties and development of QD VECSEL devices. Based on the features of QD lasers, the experimental progresses on semiconductor QD VECSELs are reviewed in detail. Besides, from the viewpoint of both single QD nanolaser and cavity quantum electrodynamics (QED), a review of a single-QD-cavity system is also presented.

## 2  Vertical-external-cavity surface-emitting lasers

The versatile semiconductor diode lasers are very widely used due to their numerous advantageous properties, such as compact size, scalability, lower integration manufacturing costs, electrical current laser excitation and modulation and excellent reliability. In this section, we focus on the basic structure, properties and advantages of semiconductor VECSELs in comparison to conventional edge-emitting lasers.

There exist two major configurations for the conventional semiconductor lasers, including edge-emitting lasers [5,14] and surface-emitting lasers [6, 11-13, 15] (Fig. 1). The edge-emitting lasers depicted in Fig. 1(a) use a waveguide to confine light to the plane of the semiconductor chip and emit elliptical light beam from the edge of the chip, the cross section of which is typically about one by several microns. The required small waveguide dimensions for single-transverse mode operation result in the asymmetric and strong angular divergence of the laser beam. The output power of edge-emitting lasers is typically limited by the required excess heat dissipation from the active region or catastrophic optical damage at the semiconductor surface [16]. Up to several hundred milliwatts of output power is achievable in a single-transverse mode waveguide configuration [15, 16]. For still wider waveguides, of the order of a 100 mm, single-stripe edge-emitting lasers can emit tens of watts of output power, but the waveguide is then highly multimoded in the plane of the chip, and output beam is very elongated with a very large, 80: 1, aspect ratio. Multiple stripe semiconductor laser bars can emit hundreds of watts, but with a highly multimoded output beam [16].

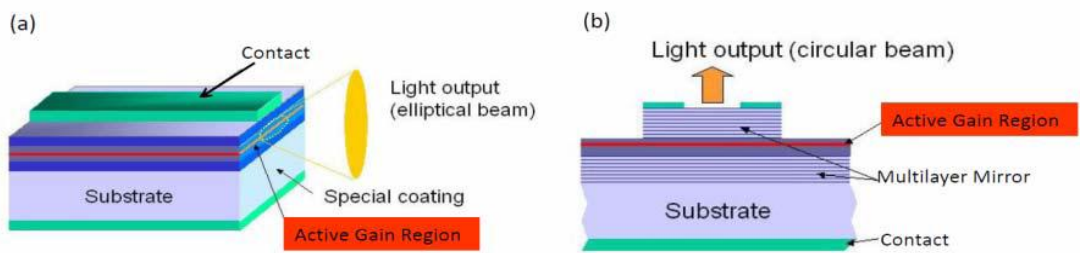

**Fig. 1** Schematic structure of semiconductor edge-emitting laser (a) and surface-emitting laser(b)

In contrast, surface-emitting lasers [11–13, 15] with laser cavity axis emit circular fundamental transverse mode light beam with powers of several milliwatts and beam diameter of several microns perpendicular to the plane of the semiconductor laser chip shown in Fig. 1(b). In particular, the output beam with circular cross section and larger beam size of VECSELs, also known as semiconductor disk lasers, has much smaller divergence than for edge-emitting lasers. Also, the required heat dissipation limits the output power and the scaling to higher powers demands larger active areas. Furthermore, semiconductor VECSELs have emerged as a low cost of fabrication, high power and high beam-quality alternative to solid-state lasers. A typical path to implement high output power laser is to use arrays of VECSELs [6, 12, 13, 16]. For much higher powers, both laser types emit highly transverse multimoded output beams.

However, high power and good beam quality cannot be achieved simultaneously with conventional semiconductor edge-emitting or surface-emitting lasers, though it is very important for many scientific and commercial laser applications. Notably, such combination is required, such as, for efficient nonlinear optical second harmonic generation [12]. It is therefore useful and important to develop a laser that exhibits simultaneously the application required and desired laser properties, such as emission wavelength, compact size, good laser beam quality and the capability of scaling up optical power to watt and higher levels with circular output beams. It is well-known that good beam quality with fundamental transverse mode operation requires strong transverse mode control of the laser cavity [11, 15]. Consequently, the VECSELs have been well developed, in which strong transverse mode control can be provided by optical cavity elements external to the laser chip, which assure that fundamental transverse mode of the laser cavity, the desired operating laser mode, has diameter approximately equal to the gain region diameter. Basic structure of VECSEL devices is depicted in Fig. 2(a) to illustrate the functions of the various layers. In a VECSEL device, a thin active semiconductor chip as the key element with typical diameters range between 50 and 500 μm, containing both a gain region and multilayer high-reflectivity mirrors, is placed on a heat sink and is excited by an incident optical pump beam. Laser cavity consists of the on chip mirror and an external spherical mirror, which defines the laser transverse mode and also serves as the output coupler. For optically pumped operation, incident pump photons with higher photon energy are absorbed in separate pump-absorbing layers, which also serve as the quantum well barriers, to generate electrons and holes. The excited carriers then diffuse to the smaller bandgap quantum wells or QD gain layers that provide gain to the optical wave, emitting lasing photons with lower photon energy. These separate pump absorption and quantum well or QD lasing emission gain layers facilitate independent optimization of the pump absorption and laser gain properties. Typically, gain layer region thickness covers several periods of this laser mode standing wave. The pump–laser photon energy difference between the incident pump photons and the emitted laser photons, due to quantum defect, together with contributions from other lasing inefficiencies, has to be dissipated as heat from the device active region. Heat dissipation from the VECSEL active semiconductor chip is provided by heat spreaders connected to heat sinks: either a soldered heat spreader below the mirror structure or a transparent heat spreader above the surface window of the chip, or possibly both. It should be pointed out that the issues of good heat dissipation and heat sinking are critical for high-power operation of all semiconductor lasers. Without these, temperature of the active region would rise and excited carriers would escape thermally from the quantum wells or QDs gain layer into the barrier region, thus depleting laser gain and turning the laser off in a thermal rollover process. Such a thermal dominant rollover mechanism typically limits the output power of VECSEL devices [17]. Fortunately, the configuration of VECSELs implements good heat sinking by placement of the transparent intracavity heat spreaders in direct contact with the laser gain region without thermally resistive laser mirrors in the path of heat dissipation [15,16,18], and have the inherent potential of producing very high powers by processing large 2-D arrays.

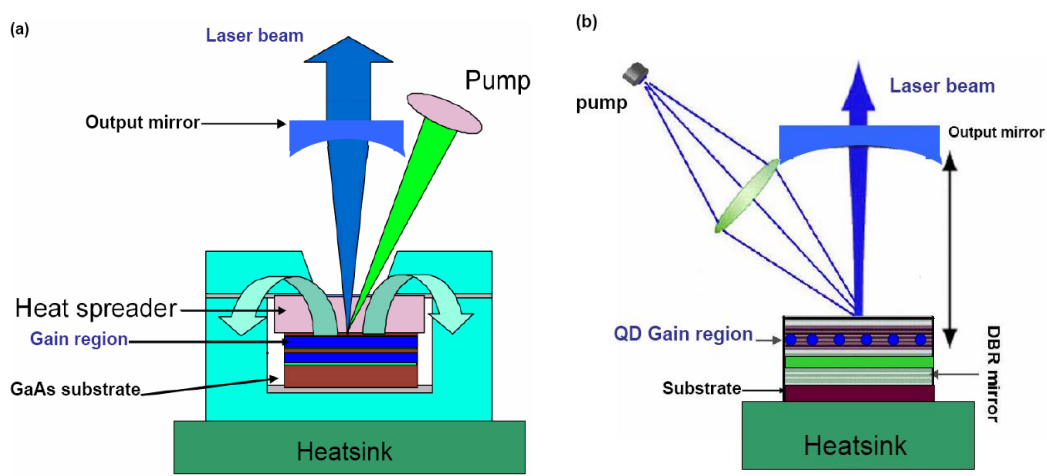

**Fig. 2** Schematic structure configuration of VECSEL device (a) and QD VECSEL device (b)

Basic configuration of VECSEL lasers enables their many key advantageous properties, which can be summarized in the following aspects:

1) Wide wavelength coverage, wavelength stability and less temperature sensitivity of wavelength: Using numerous compound semiconductor materials for different material compositions with different bandgap energies enables different lasing emission wavelengths. The use of dimensionally, or quantum-confined semiconductor active regions [6, 18] allows further control of the VECSEL emission wavelengths. By adjusting the thickness of quantum wells or diameter of quantum wires and QDs, as well as the composition of confining barrier layers, the quantum-confined electron and hole energy levels are shifted and VECSEL acquires the additional fine control of laser emission wavelength. Moreover, the lasing wavelength in a VECSEL is very stable. Note that the emission wavelength in VECSELs is ~5 times less sensitive to temperature variations than in edge-emitting lasers. The reason is that in VECSELs, the lasing wavelength is defined by the optical thickness of the singlelongitudinal- mode-cavity and that the temperature dependence of this optical thickness is minimal (the refractive index and physical thickness of the cavity have a weak dependence on temperature). On the other hand, the lasing wavelength in edge-emitting lasers is defined by the peak-gain wavelength, which has a much stronger dependence on temperature. As a consequence, the spectral linewidth for high-power arrays (where heating and temperature gradients can be significant) is much narrower in VECSEL arrays than in edge-emitting-laser arrays (bar-stacks). Note that, over a 20 ℃ change in temperature, the achieved emission wavelength in a QD VECSEL thus far can vary by less than 1.4 nm (compared to ~7 nm for edge-emitting lasers) due to the quantum effect of QDs, which we will discussed in detail later.

2) Beam Quality: VECSELs operate with a circular beam, fundamental transverse TEM00 mode, and essentially diffraction-limited low beam divergence. Several factors contribute to the beam quality in VECSELs. Most important, VECSEL device external-cavity optics defines and stabilizes the circular fundamental laser transverse mode; such optical elements and their stabilization effect are not available with the more conventional edge and surface-emitting semiconductor lasers.6 If the pump spot is too small, compared to the fundamental mode size, laser threshold will be high because of the lossy unpumped regions encountered by the laser mode. If the pump spot is too large, higher order transverse laser modes with a larger transverse extent will be excited, causing multimode laser operation and thus degraded beam quality. Optimally adjusted pump spot size gives preferentially higher gain to the fundamental laser mode, while giving excess loss from the unpumped regions to the spatially wider higher order transverse modes; this stabilizes the fundamental transverse mode operation. Using pump and laser cavity optics, VECSELs have independent control allowing matching of the pump spot size and the laser fundamental transverse mode size. Large VECSEL laser beam and pump spot sizes on the chip, tens to hundreds of microns, as compared with just a few microns for edge-emitting semiconductor lasers, contribute to the mechanical stability of the VECSEL cavity, and thus also to the stability of its fundamental transverse mode operation [6]. Through proper cavity design VECSELs can emit in a circular single transverse mode beam (circular Gaussian). This simple beam structure greatly reduces the complexity and cost of coupling/beam-shaping optics (compared with edge-emitting lasers) and increases the coupling efficiency to the fiber or pumped medium. This has been a key advantage for the VECSEL devices in low-power laser applications. Another important factor affecting spatial beam quality and stability of VECSEL devices is the negligible thermal lensing in the thin VECSEL semiconductor chip [6,19] when proper heat spreading/heat sinking is used. Thermal lensing and other beam phase profile distortions are caused by thermally induced refractive index gradients in the laser gain material. In VECSELs, thin semiconductor active region with good heat sinking implies that optical path length thermal distortions and hence beam profile changes and distortions are negligible. As a result, semiconductor VECSELs operate efficiently and with excellent beam quality across a wide range of operating power regimes from near to high above threshold.

3) Higher power per unit area: VECSELs are delivering ~1350 W/cm2 now and can deliver 2- 4 kW/cm2 in near future, while edge-emitting lasers deliver a maximum of about 500 W/cm2 because of gap between bar to bar which has to be maintained for coolant flow.

4) Scalability, heat-sinking, and packaging: Many optically pumped VECSEL devices have been reported with power levels between 10 mW and 60 W, a range of almost four orders of magnitude, while maintaining good beam quality. Such efficient power scalability is enabled by the laser mode and pump spot-size scalability on the semiconductor VECSEL chip. Since output power of semiconductor lasers is typically limited by heat dissipation and optical intensity-induced damage, increasing beam diameter in a VECSEL helps on both accounts, distributing heat and optical power over larger beam area. For well-designed heat sinking with thin semiconductor chips, heat flow from the laser active region into heat sink is essentially one-dimensional. Therefore, increasing beam area is essentially equivalent to operating multiple lasing elements in parallel, without changing thermal or optical intensity regime of the individual lasing elements. In this scenario, both output laser power and pump power scale linearly with the active area. For high-power applications, a key advantage of VECSELs is that they can be directly processed into monolithic 2-D arrays, whereas this is not possible for edge-emitting lasers. In addition, a complex and thermally inefficient mounting scheme is required to mount edge-emitter laser bars in stacks. Good heat-sinking and packaging of VECSELs have been two key advantages. Due to the simple processing and heat-sinking technology it is much easier to package 2-D VECSEL arrays than an equivalent edge-emitting laser bar-stack. Mounting of large high-power VECSEL 2-D arrays in a "junction-down" configuration is straightforward, making the heat-removal process very efficient. In fact, VECSEL manufacturing follows the standard, well-established IC silicon industry processing.

5) High temperature operation: VCSEL devices can be operated without refrigerationbecause they can be operated at temperatures to 80°C, The cooling system becomes very small, rugged and portable with this approach.

6) Functional versatility through intracavity optical elements: Although the external optical cavity, which controls the laser transverse modes, would make these lasers more complex and requiring assembly and alignment as compared with the simple integrated surfaceemitting lasers, such an external cavity of VECSELs gives tremendous versatility to VECSEL device configurations and functions. Flexible design and construction of VECSEL device cavities [20-24], such as linear two-mirror cavity, three-mirror V-shaped cavity, and four-mirror Z-shaped cavity, allow flexible insertion of intracavity optical elements. Such intracavity functional elements are very difficult to use with integrated semiconductor devices. One important option enabled by the external cavity is the insertion of intracavity spectral filters, such as Brewster's angle birefringent filters [25, 26], volume gratings [27], or high-reflectivity gratings [28], to control longitudinal spectral modes of the laser and possibly to select a single longitudinal lasing mode. Moreover, the insertion of intracavity saturable absorber elements to external cavity enables VECSEL to achieve laser passive mode locking with picosecond and subpicosecond pulse generation. In this case, the length of the external short cavity allows control of the pulse repetition rates, with rates as high as 50GHz [29, 30]. External cavity optics also allows different beam spot sizes on the gain and absorber elements required to achieve mode locking, which controls optical intensity of the beam spots [20, 31]. The open cavity of VECSELs implements placement of transparent intracavity heat spreaders in direct contact with the laser gain element without thermally resistive laser mirrors in the path of heat dissipation [24, 32, 33]. Since thermal management is critical for high-power VECSEL operation, the utilization of such heat spreaders tremendously broadens laser design options with chip gain, mirror, and substrate materials that do not allow effective heat removal through the on-chip mirror. Another option for VECSELs allowed by the external cavity is the microchip laser regime [6, 34, 35]. Low intracavity loss of VECSELs, combined with their wide gain bandwidth, allows insertion of intracavity absorption cells, such as gas cells, for intracavity laser absorption spectroscopy (ICLAS) performing sensitive measurements of extremely weak absorption lines [36]. Besides, availability of such high intracavity power, together with high beam quality, allows very efficient nonlinear optical operation, such as second harmonic generation, by inserting nonlinear optical crystals inside the external laser cavity [19, 24, 26]. Using intracavity second harmonic generation, VECSELs have provided efficient laser output at wavelengths not accessible by other laser materials and techniques.

To summarize, these combination advantageous properties of semiconductor VECSEL devices with excellent beam quality operate efficiently across a wide range of operating power regimes from milliwatts to tens of watts with single or multiple semiconductor gain chips resulting in versatile laser applications. In their early stage application, VECSEL devices were developed as high-power single-mode fiber-coupled sources at 980nm for pumping Er-doped fiber and glass-waveguide amplifiers for optical fiber telecommunications systems [6]. Commercial 850 nm GaAs VCSELs have been well established for these short-reach applications. Gigabit Ethernet and Fiber Channel are currently major markets for VECSELs [6, 13]. Long wavelength QD VECSELs emitting at 1.2–1.33 μm are currently attracting much interest for use in single-mode fiber metropolitan area and wide-area networks. Further optimization of the quantum defect and electron confinement energy is required for room temperature operation of high-performance QD VECSEL devices, the structure of which is depicted in Fig. 2(b). Note that in practical application there is some trade-off between output power and beam quality in VECSELs: a multimode laser beam can better overlap the pump spot and thus produce somewhat higher output power [13, 16, 37].

## 3   Quantum dot laser

The approach of quantum effects aimed at modifying the electronic density of states in active layers of a heterostructure semiconductor laser to achieve a spectral tuning ability without changing chemical composition [38, 39]. In this section, we focus on the basic properties and development of semiconductor QD lasers.

As shown in Fig. 3, the quantum effect increases with the progressive reduction in dimensionality: from the square root dependence of a bulk solid, through a step-like dependence of quantum wells and an inverse square root dependence of quantum wires to the discrete delta function of QDs. As a result, nonequilibrium charge carriers injected into a size-quantized semiconductor gradually concentrate on the band edges. Note that, in bulk material charge carriers are redistributed within the continuous distribution function to higher energy states as the temperature is increased. These carriers do not contribute to the inversion at lasing energy, and thus additional injection current is required to maintain a constant inversion condition. The effect of QD optical gain was analyzed for laser structures based on InP and GaAs QD materials [40-43]. Theoretical modeling work assumed a single active layer of QDs, a cubic shape of uniform size in the QD ensemble, and hence an equal localization of carriers in the QDs. The relationship for active GaAs layers embedded in $Al_{0.2}Ga_{0.8}As$ shows a clear increase in the maximum gain as the dimensionality is reduced. The increase in maximum gain is also reflected in a decrease in the lasing threshold current density $J_{th}$. The lasing threshold is generally reached, if the modal gain $g_{mod}$ just equilibrates the internal losses $\alpha_{int}$ and those of the cavity mirrors $\alpha_{mirr}$,

$$g_{mod} = \alpha_{int} + \alpha_{mirr}. \quad (1)$$

Notably, QDs are nanosized semiconductors in the range of the de Broglie wavelength of charge carriers in all spatial directions, leading to atom-like, fully quantized states of confined electrons and holes. The discrete nature of the energy states causes the unique properties of QDs, not observed with higher dimensionality systems. Although the optical confinement factor of a QD layer is significantly smaller than that of a QW, the material gain of QD media $g_{mat}$ exceeds that of QWs by far [6,38], because of the increase in the density of states described above, as well as an increased overlap of confined electron and hole wave functions and thereby an increase in the exciton binding energy due to quantum confinement. The oscillator strength of radiative recombination is thereby significantly enhanced. Device applications of carrier confinement of QD media in all three dimensions promised significant advantages over the one-dimensional confinement of quantum wells[39,40]. Since the number of states in a QD layer is small compared to those of a QW, a low lasing threshold and a low gain saturation level are easily reached in single active layers for QD lasers. In order to increase both confinement factor and gain saturation level, actual QD lasers generally comprise a stack of QD layers. As a result of confinement factor and material gain, a largely comparable modal gain is found in QD lasers. Moreover, the behavior of decrease in the temperature sensitivity of the threshold current density can be expected and observed due to the reduction in dimensionality, leading in the ultimate limit of a zero-dimensional nanostructure to a lasing threshold current density Jth being independent of temperature. From the viewpoint of low-dimensional condensed matter physics, low temperature sensitivity of zero-dimensional nanostructures is a direct consequence of the modified electronic density of states [38, 39]. As shown in Fig. 3, the occupation in a QD is described by a delta function. The charge carriers cannot be redistributed if excited states lie sufficiently above the ground state. Besides, it is worthwhile to emphasize that quantum size effects of QDs leading to an increase in the density of states near the band edges can lower the lasing threshold [38]. Therefore, an ultimate microscopic limit of thresholdless semiconductor lasers is QD laser. Besides, a QD laser is expected to work under the thresholdless operation only if the exciton resonance wavelength is the same as that of the resonant wavelength of the nanocavity, and when the position is located where the electric field of the optical cavity is strongest.

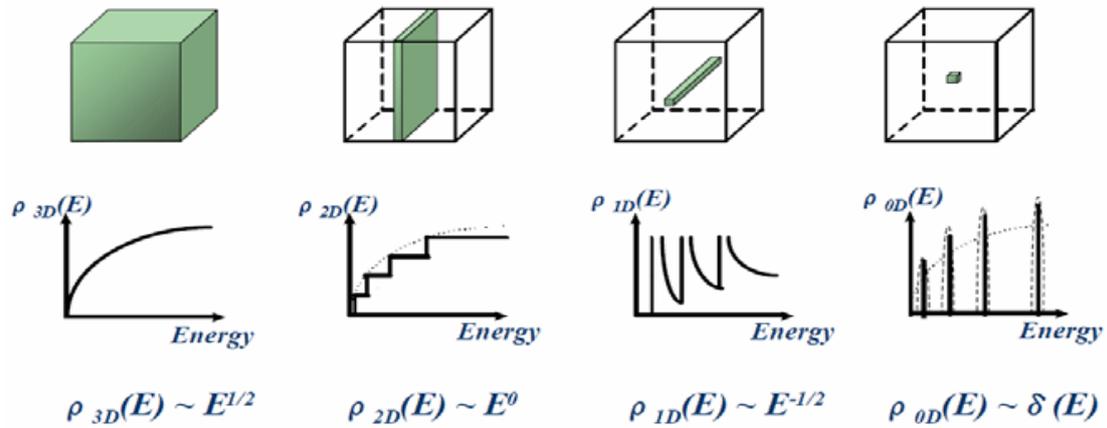

**Fig. 3** Schematic illustration of quantum effects (density of states) from bulk, quantum well, quantum wire to QD

As a matter of fact, research work on low dimensional nanostructures as gain media for semiconductor lasers first focused on quantum wells due to lacking reliable methods for fabricating sufficiently high-quality QDs at that time [6,31,36]. However semiconductor quantum well lasers suffer from limitations such as relatively high threshold powers, multimode operation, problems with direct modulation above 20GHz, and difficulty in growing their Distributed Bragg Reflectors (DBR) for long-haul communications wavelengths. Moreover, the discrete nature of the energy states or electronic density of states causes the emergence of unique properties of QDs, which cannot be observed in higher dimensionality nanostructures [39]. As a consequence, the threshold of QD lasers is ultimately minimized (the so-called thresholdless laser) by eliminating spontaneous emission as much as possible [39, 40]. Early attempts to implement the QDs by dry etching of quantum well structures into a laser resulted in devices with poor performance [31]. In the 1990s, the efficient approach of self-organized Stranski–Krastanow growth was applied to realize ensembles of QDs with a high areal density [40]. Since then, much work was devoted to exploit the ability of high quality Stranski–Krastanow QDs to emit at wavelength from 650nm to 2.8 μm [45–70]. It should be noted that the radiative recombination was strongly limited due to the relaxation of captured carriers hampered by a phonon bottleneck effect, and the orthogonality of electron and hole wave functions. Besides, the inhomogeneous broadening within a QD ensemble and the assumed small total number of QDs participating in lasing operation were considered to be a major drawback to achieve sufficient output power [43, 54, 55]. The first injection lasers based on self-organized QDs grown using MBE were edge-emitting ridge-waveguide devices with a single layer of In0.5Ga0.5As/GaAs Stranski–Krastanow QDs [56]. The devices demonstrated the predicted advantages [41, 43, 54–56] of zero-dimensional nanostructures over those of higher dimensionality. A low threshold current density of 120 Acm-2 with a low temperature sensitivity expressed by a high characteristic temperature of 350 K was observed at lowtemperature operation. However, the excellent performance was maintained only up to 120 K as a result of several processes, such as thermal redistribution of carriers to nonlasing states within the QDs, thermally induced escape of carriers out of the QDs, and Auger nonradiative mechanisms [57]. Later, various approaches were employed to achieve improvements of laser characteristics by MOVPE. The first MOVPE-grown QD lasers used either an InAs QD stack [58] or an In0.5Ga0.5As QD stack [59]. Both approaches achieved ground-state lasing at room temperature, featuring a critical temperature of 385K for the 10-fold In0.5Ga0.5As QD stack laser. The stacks were grown at the same low temperature for QD layers and GaAs spacers. The performance has further been substantially improved by introducing temperature cycling to grow the spacer layers at increased temperature and to smoothen the spacer-QD interfaces [60]. QD layers represent small grouped peaks in the gain region, which is clad by the DBR and a window layer on the left- and right-hand sides, respectively. QD lasers benefit from the reduced lateral charge carrier diffusion due to trapping in the QDs [61-63]. The effect was shown to suppress beam filamentation observed in quantum well lasers [62]. Moreover, nonradiative surface recombination decreases. As a consequence, the robustness against facet degradation in high-power operation is enhanced, and deep etch through the active region in narrow-stripe ridges is possible without surface passivation [63]. Furthermore, using the Dot-in-a-Well (DWELL) approach through inserting a strain-reducing layer to reduce the strain exerted on the buried QD layer with respect to a cap and carefully adjusting the growth parameters for each individual QD layer in the stack to avoid defect formation, lasing at 1.25μm with a very low threshold current density of 66 Acm-2 and 94% internal quantum efficiency was obtained [64]. Besides, the target of lasing at 1.35 μm from ten stacked InAs-Sb:GaAs QD layers grown using MOVPE was accomplished with QWELL structure devices [65].

## 4 Vertical-external-cavity surface-emitting lasers based on quantum dots

4.1 Introduction to QD VECSELs

In this section, we focus on the properties and recent developments of semiconductor QD VECSELs based on two typical alternative paths for high-density QDs of gain layers. The unique properties of QDs inspired intensive studies to implement them into semiconductor VECSEL devices, as depicted in Fig. 2(b). QD VECSELs as a relatively recent type of semiconductor lasers are key building elements for optical communication systems. Historic trend predicts fourfold increase in the maximum commercial single channel data rate every five years [66]. At present, QD VECSELs span the 650 nm to 1.8 μm spectral range, with record cw output powers at room temperature [6, 38, 66]. QD VECSELs are usually optically pumped using diode lasers, with a cavity formed by an epitaxially grown gain mirror, consisting of an active region on top a distributed Bragg reflector (DBR), and a second external mirror which also acts as an output coupler. The QD VECSEL allows the addition of intracavity elements including nonlinear crystals for frequency downconversion. These techniques have resulted in the use of near-IR VECSELs as frequency-doubled and quadrupled sources for visible and UV wavelengths. QDs have to be placed near the antinodes of this standing wave in order to provide efficient gain to the laser. This is the socalled resonant periodic gain (RPG) arrangement [6, 13, 42] of placing QD gain layer near a given standing wave antinode. In addition, QD VECSEL devices also demonstrate excellent dynamic performances such as low threshold currents (a few micro-amps), low noise operation and high-speed digital modulation (10 Gb/s) [66, 67].

The crucial requirement of a high QD density in the volume of the gain medium can be realized by a high areal density and a dense stacking of QD layers technically, which is particularly challenging for QDs emitting at long wavelength because these QDs are comparably large and have consequently large strain fields. The design and growth procedure must avoid a strain-induced formation of defects, which may degrade the performance of the device. In addition, the strain-induced structural coupling of QDs in adjacent layers may lead to a vertical alignment with an increased inhomogeneous broadening due to a gradually increasing QD size [68]. Based on the thermal stability of covered dots below 600°C, a procedure to flatten the growth front prior to QD layer deposition and to overgrow the dots was established that maintains a high radiative recombination efficiency of the QDs [69]. One strategy that could be used to get to longer wavelengths would be to manage the strain through employing strain compensation; however, this technique has its limitations. The use of RPG structures with a single-QD layer per antinode is another way to control strain in such lasers. Since 2008, several first realized QD gain chips that implemented a layout of RPG structure to selectively enhance gain at the operating wavelength have further been reported [13, 71-74].

Advances in growth control of high-quality QD layers eventually allowed the implementation of QDs also in gain media of semiconductor VECSELs [70–76]. Consequently, QDs for four working wavelengths of 940, 1040, 1180, and 1250 nm have been successfully implemented into the resonant gain structures of QD VECSEL devices. The Stranski–Krastanow and the submonolayer growth modes as two typical alternative paths for high-density QDs have been used for maximization of the gain region so far, and therefore the operation of VECSEL devices have been demonstrated with both the Stranski–Krastanow QDs [71,72,74] and submonolayer QDs [73,75,76]. Both kinds of QDs have quite different optical properties. The Stranski–Krastanow QDs show a very broad spectrum. On the other hand, the submonolayers QDs exhibit a narrow luminescence with a significant thermal shift. Careful adjustment of emission wavelength is therefore required. The first design intends to maximize the modal gain per dot layer, while the second design aims for a maximum total modal gain by increasing the number of QD layers.

4.2 VECSELs with Stranski-Krastanow QDs

The first implementation of Stranski–Krastanow QDs employed in semiconductor QD VECSELs for 1040nm emission [71], in which the QD layers were integrated into a resonant gain structure. Surface PL occurs at the wavelength of the small dip in the reflectivity of the DBR stopband that is attributed to the subcavity resonance. The lasing spectrum of the Stranski-Krastanow-QD VECSEL device in Ref. [71] shows Fabry-Perot fringes due to the etalon effect of the diamond intra cavity heat spreader. The threshold pump power is 6.5 W and the slope efficiency is 6.7%. At high pump power, an onset of thermal rollover occurs. As we have mentioned before, one potential benefit of employing QDs instead of quantum wells in a gain medium is the achievement of a low lasing threshold with much smaller temperature dependence. Moreover, by way of elastic strain relaxation in a uniform arrangement of 7×3 Stranski-Krastanow grown QD layers, GaAs QD laser devices for 1220 nm operation have been achieved to extend the operation wavelength to the near-IR range [72]. The device showed a threshold pump power of 0.48 W at 15°C for a high reflectivity (99.8%) output coupler mirror. Notably, the results demonstrate temperature-stable operation in the measured range. The dependence on pump power density is only 0.027 nm (kWcm-2)-1. The center wavelength of the emission shifts only by 0.06nmK-1, leaving the output power largely unaffected. Such shift is almost an order of magnitude smaller than typical values of 0.3nmK-1 observed for VECSELs based on GaInNAs quantum wells [67], and the temperature-independent differential efficiency of the device is about 2%. The use of InAs QD active regions for VECSELs could achieve high-power lasers in the 900–1300 nm wavelength range.

Very recently, Albrecht et al. [13] have demonstrate the growth of 1250 nm emission wavelength InAs-QD VECSELs and achieved better lasing performance by designing two different RPG structures with increased spacing between the QD layers, thus allowing the crystal to recover between each QD layer growth. The first laser is based on the traditional multi-QD layer per antinode design, which is referred to as the "4×3" structure (four sets of three QD layers). The second laser that utilize a single-QD layer per antinode with "12×1" structure has demonstrated 3.25 W of cw output power using a CVD diamond heat spreader mounted to the thinned substrate of the 12×1 structure. The active region growth for the two VECSELs is based on identical QD structures, and the emission wavelength for the two structures is around 1250 nm. For both designs, highly optimized QD structures for maximum gain are made through the use of DWELL structures. The QDs are designed for 1250 nm emission and are grown with a total thickness of 1.68 monolayers of InAs deposited after growing 1 nm of 7 nm total thickness In0.15Ga0.85As quantum well. The only variation in the two VECSELs is the design of the RPG active region and the active RPG region is grown at a substrate temperature of 480°C. The increased separation between the dot layer in the 12×1 structure compared to the 4×3 resulted in superior lasing performance with regard to the threshold pump-power density, the slope efficiency, and the maximum output power. An important issue for the further development of QD VECSELs is the dense stacking of QD layers.

4.3 VECSELs with submonolayer QDs

As a practical alternative approach, the submonolayer QDs were implemented in semiconductor QD disk lasers for 940 nm [75] and 1040 nm emission [73]. The submonolayer QDs were grown by cycled depositions of pure binaries InAs and GaAs. The tuning of the emission wavelength may be performed by adjusting either the InAs deposition duration or the number of InAs/GaAs cycles. As a matter of fact, the wavelength control is somewhat facilitated by adjusting the number of cycles, yielding robust reproducibility with similar optical performance. The wavelength tuning of submonolayer QDs must consider proper alignment of the PL peak wavelength with the cavity resonance of the VECSEL structure. The submonolayer QDs for the emission wavelength of 940 nm were grown using a fivefold cycle of 0.5 monolayer InAs and 2.3 monolayer GaAs. Studies of stacking three such submonolayer QD layers using GaAs spacers from 60 to 10 nm showed a constant high PL intensity down to 20 nm thickness and a drop of intensity for thinner spacers. A thickness of 20 nm for three layers in an antinode was therefore chosen for device applications to obtain a large overlap to the optical field in the gain region.

Submonolayer QDs for 1040 nm operation were grown in the same way with 10 InAs/GaAs cycles of nominally 0.5 monolayer InAs and 2.3 monolayer GaAs as those for 940nm emission [73]. A minimum spacer thickness of 20 nm could be achieved without structural degradation. Wavelength tuning was performed by increasing the number of submonolayer cycles per submonolayer-QD layer. Limits for the given conditions were found to be 1070 and 960 nm for the long and short wavelength side, respectively. Results on QD VECSELs achieved thus far yield a higher gain achieved with submonolayer QDs, accompanied by a smaller wavelength range of gain. For the case of 1035nm wavelength emission, 13 submonolayer-QD layers, each comprising 10-fold cycled depositions of nominally 0.5 monolayer InAs and 2.3 monolayer GaAs, were integrated into a gain structure [67]. Compared to Stranski–Krastanow QDs, the larger gain of submonolayer-QDs leads to increased slope efficiency. Characteristics of the device yield a slope efficiency of up to 12.4%. Due to a larger efficiency, no significant thermal rollover is observed. A maximum cw output power of 1.4W is obtained for 1% outcoupling. Good characteristics were likewise obtained from a submonolayer-QD VECSEL fabricated for 940 nm emission [75]. An output power of 0.5W at 1% outcoupling was achieved using 10 layers of fivefold cycled submonolayer-QDs. Notably, the low value originates from the limited modal gain of the structure and is an issue of future improvements.

## 5  Ultimate nanolasers: single Quantum Dot VECSEL devices

In this part we discuss the development and current status of a single QD-cavity coupled system. Over the past ten years, the possibility of realizing a single QD nanolaser (the socalled thresholdless laser) in which the threshold is ultimately eliminating has been drawing intensive attention. As an ultimate microscopic limit of thresholdless semiconductor lasers, single-QD nanolaser, i.e., a single QD coupled to a single mode of an optical cavity shown in Fig. 4, allow us to investigate the light–matter interaction down to the single particle level entering the so-called cavity quantum electrodynamics (QED) regime [4,77,78]. As early as in 1999, Pelton et al. [77] firstly proposed a single QD microcavity system as a novel ultralow threshold laser device, consisting of a single InAs/GaAs self-assembled QD coupled to a highfinesse microsphere cavity. Later, S. Strauf and co-workers have demonstrated that very few (2–4) QDs as a gain medium are sufficient to realize a photonic crystal (PhC) laser based on a high-quality nanocavity [78]. Photon correlation measurements show a transition from a thermal to a coherent light state proving that lasing action occurs at ultralow thresholds. They conclude that the quasicontinuous QD states become crucial since they provide an energytransfer channel into the lasing mode, effectively leading to a self-tuned resonance for the gain medium. In 2006, for applications in low THz sources (1 THz) for molecular detection in chemistry and biology as well as THz-band communications between 0.8–3 THz, we proposed and examined a terahertz laser model based on a single InAs semiconductor QD emitter in a microcavity [39], in which single InAs QD is selectively placed in a high quality microdisk cavity, as shown in Figs. 4(a) and 4(b), which is resonant with an intersublevel transition (ISBT) between quasicontinuous states of QD. Specifically, to achieve the desired long-wavelength of THz, an intersublevel lasing transition is from an excited conduction band state to a lower conduction band state between quasicontinuous states of the InAs QD emitters. The microdisk cavity is tuned to the radiative transition resonance, so likely the final transition emptying this state will be off resonance [39]. An increase in DOS of the lasing mode causes significant enhancement of the spontaneous emission rate (Purcell effect) [77]. This consequently enables larger fractions of photons to be emitted into the lasing mode with respect to all other modes (denoted as spontaneous emission-coupling factor *β*) and reduces the lasing threshold [39, 77, 78, 82, 83]. Here we want to point out that from the viewpoint of photonic quantum information processing, such a single-QD-microcavity system can also be well-suited for single-photon generation [1, 4, 78, 79].

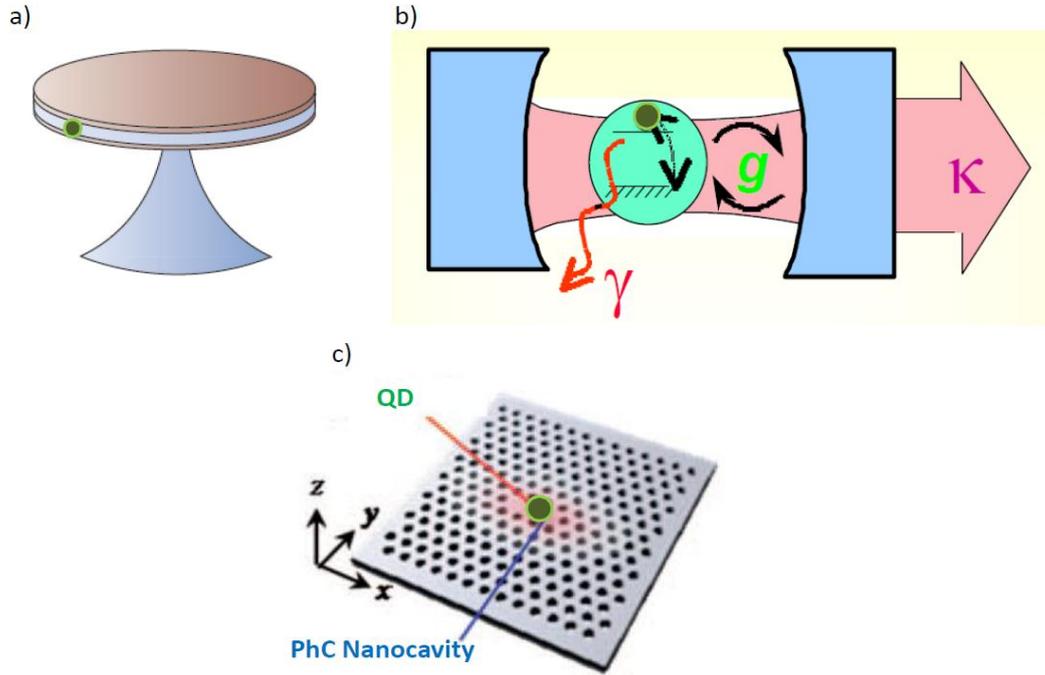

**Fig. 4** Schematic structure (a) and corresponding working mechanism (b) of single QD microdisk nanolaser, in which an atomic QD in a small-volume microdisk cavity with dephasing rate *γ* coupled to a cavity with photon loss rate *κ* by coupling strength *g*. c) Schematic illustration of the QD-PhC-nanocavity system considered, in which a single QD embedded in a PhC nanocavity

Furthermore, PhC nanocavities [81, 84–89] are also promising candidates for the trapping of light in ultrasmall volumes with high Q−factor. And a QD with 3D carrier confinement as the most suitable gain medium can be introduced into a PhC nanocavity, avoiding the effects of nonradiative centers that are inevitably introduced in the fabrication of PhCs and thereby restricting the coupling of the QD gain medium to the lasing mode. Recently, such a coupled system consisting of a nanocavity and a QD shown in Fig. 4(c) has been further extensively investigated because of its promising applications such as quantum information processing [4,83–86], single photon sources [1,81,85], and ultimately low-threshold nanolasers [78,80,86–89]. For such a single QD PhC nanocavity system, utilization of a single mode cavity with a sufficiently high Q factor as a lasing mode, the modal volume of which should be as small as possible to maximize the interaction with the single QD gain medium, is key to realize a thresholdless QD nanolaser. For such a single-QD-nanocavity system the two phenomena that light should be emitted only at the QD transition energy under off-resonant conditions and vacuum Rabi splitting (VRS) should occur under on-resonant conditions are expected to occur naturally, analogous to the atomic systems in cavity QED [89]. However, experimental reports have shown clear deviations from these features. First, light emission from the cavity occurs even when it is largely detuned from the QD [83, 84, 86, 87]. Second, spectral triplets are formed by additional bare cavity lines between the VRS lines under on-resonant conditions [86, 87]. These features are unique to semiconductor nanostructure systems and are not predicted by conventional cavity QED in atomic systems. In our study the QD was typically modeled by a simple atomic two-level system and the effect of pure dephasing was to broaden the transition energy of the two-level system [39,86], which makes it possible for the tail of the transition energy to overlap with the cavity energy, and generates an additional way in which to interact with the cavity. Owing to this interaction, a cavity photon is created with the transition of the excited two-level system to the ground state, a process generally known as the anti-Zeno effect (AZE), induced by pure dephasing in the QD. In most cases, this effect could be negligible, because the tail of the atomic two-level system transition energy at the cavity energy is so weak that the interaction is not large enough to overcome the direct spontaneous emission from the two-level system to free space. However, the unique situation achieved in nanocavity QED systems strongly enhances the AZE, resulting in off-resonant cavity light emission due to the following reasons. First, the coupling constant between the cavity and the two-level system is large due to the small cavity volume. Second, the quality factor of the cavity is relatively large. Third, direct spontaneous emission from the two-level system to free space is strongly suppressed due to the in-plane photonic band-gap effect in the case of the two-dimensional PhC nanocavities shown in Fig. 4(c) [84–86]. These features are well described by a factor $F$, which is given by

$$F \approx \frac{\Gamma_{spon} + \gamma_{phase}}{\Gamma_{spon}(\delta\omega_{TLS,c}/g_{TLS,c})^2 + \Gamma_c + (\Gamma_{spon} + \gamma_{phase})}, \quad (2)$$

where $2\Gamma_{spon}$ is the direct spontaneous emission rate of the TLS to free space, $2\Gamma_c$ is the damping rate of the cavity, which is determined by its $Q$ factor, $2\gamma_{phase}$ is the pure dephasing rate of the TLS, $\delta\omega_{TLS,c}$ is the detuning of the TLS from the cavity, and $g_{TLS,c}$ is the coupling constant between the TLS and the cavity. Thus, the off-resonant cavity light emission can successfully be explained by this nanocavity-enhanced AZE, and we conclude that the phenomenon of most low-dimensional semiconductor nanostructures in PhC nanocavities cannot be fully describes by atomic two-level systems in cavity QED. Furthermore, by comparing the intensity of the central peak in the triplet with the values predicted by the factor F, we conclude that the on-resonant triplet is explained well by the AZE from the detuned states under a low-excitation regime. A similar analysis was also reported independently by Yamaguchi *et al.* [85]. Their calculated results reproduce the features of the experimentally observed on-resonant spectral triplet, the origin of which has puzzled the community for several years. Such a formalism should prove to be useful for future studies, such as practical analyses of quantum information processing using QD spins and cavity QED [4,80, 85-87], as well as studies of the design and optimization of the performance of single photon sources [1,82,85]. We believe that these results above will stimulate a full understanding of the physical nature of solid-state cavity QED systems. A pulsed QD nanocavity could serve as a single photon turnstile that is more deterministic than one based on Purcell enhancement [77, 78, 80]. Controlled quantum entanglement in a QD nanocavity could be used for quantum state transfer where true strong-coupling serves as a bidirectional interface between semiconductor and photonic quantum states, as required in a quantum information network [1, 90]. In the future, large-scale monolithic photonic integrated circuits (MPhICs) represent a significant technology innovation that simplifies optical communication system design, reduces space and power consumption, and improves reliability. In addition, by lowering the cost of optical-to-electrical-to-optical (OEO) conversion in optical networks, QD nanolaser provides a transformational opportunity to embrace the use of monolithic electronic silicon ICs and system software in a "digital" optical network to maximize system functionality, improve service flexibility, and simplify network operations.

## 6  Conclusions

This review has discussed properties and advantages of semiconductor VECSELs, and properties and recent development QD lasers, as a building block for practical optical communication systems. In particular, this review outlines the properties and recent proofs of developments of QD VECSEL devices in detail. In the final section of this review, from the viewpoint of both the single QD nanolaser device and cavity QED, we presented the theory, model and analysis of a coupled system consisting of a nanocavity and a QD. Although recent pioneering developments in nanofabrication technologies are beginning to realize such systems, precise tuning between the single QD and PhC nanocavity which is critical to the realization of a single QD nanolaser and single-photon source needs to be further improved in future. The presented results in this review are important for the future development of QD VECSEL technology or even MPhICs and quantum information devices.

**Acknowledgments** The research conducted in State Key Laboratory of Functional Material for Informatics has been supported by the Major State Basic Research Development Program of China (Grant No. 2009CB930600), and by a Strategic Research Grant (No. 7008101) from City University of Hong Kong.